\documentclass[traditabstract]{aa} 
 \usepackage{longtable}
 \usepackage{epsfig}
\usepackage{xcolor}
\usepackage{txfonts}
\usepackage{amssymb}
\usepackage{natbib}       
\usepackage{url}       

\bibpunct{(}{)}{;}{a}{}{,}
\newcommand{\be}{\begin{equation}}
\newcommand{\ee}{\end{equation}}
\newcommand{\bea}{\begin{eqnarray}}
\newcommand{\eea}{\end{eqnarray}}

\newcommand{\vv}{\rm v}

\begin{document}

\title{X-ray spectroscopy method of white dwarf mass determination in intermediate polars. External systematic uncertainties.}
\author{
V.~F. Suleimanov\inst{1}
\and
L. Ducci\inst{1}
\and
V. Doroshenko\inst{1}
\and
K. Werner\inst{1}}

\institute{
Institut f\"ur Astronomie und Astrophysik, Kepler Center for Astro and
Particle Physics, Universit\"at T\"ubingen, Sand 1, 72076 T\"ubingen, Germany\\ \email{suleimanov@astro.uni-tuebingen.de}
}

\date{Received xxx / Accepted xxx}

   \authorrunning{Suleimanov et al.}
   \titlerunning{White dwarf masses in IPs. External uncertainties.}

\abstract
{The masses of white dwarfs (WDs) in intermediate polars (IPs) can be determined from the shape of their hard X-ray spectra. Here we study the importance of all possible systematic uncertainties in this  X-ray spectroscopy method, including finite radii and rotation of magnetospheres, finite accretion column height, and accretion-flow inclination relative to the WD surface. We also investigate the importance of accretion-heated envelopes on WD surfaces in IPs that are increasing WD radii. Their presence changes the commonly used mass-radius relation for cold WDs. As a first approximation we use thick ($10^{-4}M_\odot$) hydrogen envelope models with a surface temperature of 30\,kK. We present a new model grid of hard X-ray spectra of high-luminous IPs computed  using a new mass-radius relation. This grid is used for fitting Swift/BAT spectra of 47 IPs. The average WD mass in this sample is 0.82\,$M_\odot$ and coincides with the average WD mass in cataclysmic variables obtained by optical methods. This means that the calculated hard X-ray spectra and the assumptions made that the magnetospheric radii in IPs are close to the corotation radii and that the relative heights of the accretion columns are small are basically correct, because most IPs have high luminosities. However, this universal grid (as well as previous universal grids) cannot give correct results for the low-luminous IPs with probably relatively tall accretion columns on the WD surfaces. Such IPs have to be investigated with individual accretion column models.} 

\maketitle
%


\section{Introduction}

Cataclysmic variables (CVs) are close binary systems with a white dwarf (WD) as a primary and a normal star overfilling its Roche lobe as a secondary \citep[see detail description in][]{Warn:03}. 
One of the challenges associated with these accreting binary systems is understanding why the masses of WDs in CVs differ from the ones of single WDs \citep[see, e.g.,][]{ZS20}. The average mass of WDs in CVs is about 0.8\,$M_\odot$ \citep{Zorotovic.etal:11}, and this value is significantly higher than the average mass of isolated WDs, 0.6$-$0.7\,$M_\odot$ \citep{2007MNRAS.375.1315K}. One possible reason is a faster evolution of the systems with low-mass WDs due to more effective angular momentum loss during nova explosions \citep{2016MNRAS.455L..16S}. Nevertheless, the accurate WD mass determination in CVs is still a topical problem.

White dwarf masses in magnetized CVs, intermediate polars (IPs) and, in some cases, in polars (AM Her type CVs) can be determined using hard X-ray spectroscopy. Accreted matter forms an accretion disk in IPs, which is, however, destroyed by the WD magnetic field near the magnetospheric radius, $R_{\rm m}$. Subsequently, the matter flows along the magnetic field lines and forms strong hydrodynamical shocks and hot optically thin post-shock structures (PSRs) near the magnetic poles of the WD \citep{1973PThPh..49.1184A, 1976MNRAS.175...43F}. The temperatures of PSRs are close to virial temperatures at the WD surface, reaching several tens of keV. As a result, IPs are sources of hard X-ray radiation with spectra close to one-temperature bremsstrahlung \citep[see e.g.,][]{Mukai:17, deMartino.etal:20, 2020NewAR..9101547L}. Model spectra of PSRs depend mainly on the WD mass, and this opens the possibility to determine masses \citep[see the first application by][]{1981ApJ...250..723R}. 

Recently, WD masses in significant numbers of IPs were determined using NuSTAR and Swift/BAT observations \citep[][see detailed description of the method in the next section]{Suleimanov.etal:19, Shaw.etal:20}. The average WD mass obtained in these works is again about 0.8\,$M_\odot$ \citep[0.79$\pm 0.16$\,M$_\odot$ from 35 IPs and 0.77$\pm 0.12$\,M$_\odot$ from 17 IPs plus two polars, respectively;][]{Suleimanov.etal:19,Shaw.etal:20}. A comparison of this value with the average WD mass in CVs obtained with optical methods by \citet{Zorotovic.etal:11} (0.82$\pm 0.15\,M_\odot$ from 32 CVs) and \citet{2022MNRAS.510.6110P} ($0.81^{+0.16}_{-0.20}\,M_\odot$ from 89 CVs) shows that they closely coincide within the errors, but the average WD mass obtained from X-ray observations is slightly lower by 0.02$-$0.05\,$M_\odot$. We note that the average WD mass in CVs, determined from the ratio of fluxes in emission lines Fe\,XXVI-Ly$\alpha$ to Fe\,XXV-He$\alpha$ \citep[58 CVs, 0.81$\pm 0.21\,M_\odot$,][]{2022RAA....22d5003Y}, is also consistent with the results presented above.
 
However, recently a few works have been published  in which the WD mass in three IPs was measured from optical observations, namely in GK\,Per \citep{2021MNRAS.507.5805A}, XY\,Ari \citep{2023arXiv230408524A}, and DO Dra \citep{2024arXiv240717562A}. 
These masses were obtained without any additional assumptions, such as a  dependence between the orbital period and the donor star mass \citep[see, e.g.,][]{2011ApJS..194...28K}, which 
was often used before.
They are significantly higher than the masses found using X-ray spectroscopy and in previous determinations obtained with other methods (see Table~\ref{tab:mass}). The differences reach 0.15$-$0.24\,$M_\odot$, but formally they do not exceed three standard deviations (3 $\sigma$). On this basis, the authors of these papers express doubts about the reliability of the WD masses obtained by X-ray spectroscopy in the IPs as a whole. In this paper, we aim to find out what unaccounted systematic uncertainties could lead to such differences and to understand whether it is possible to trust the WD masses obtained by X-ray spectroscopy for other IPs.

We also took into consideration the WD mass values in the IP EX Hya, obtained by  \citet{Suleimanov.etal:19} and using optical methods 
\citep{2016MNRAS.461.1576E, BR:08}. The WD mass values of EX Hya obtained by optical methods can also be considered reliable, since the system is eclipsing and the inclination angle is well known. We do not provide all the WD mass estimations in this system made over the past 45 years \citep[see detailed review in][]{Echevarria.etal:16}, but only the most recent ones, which we consider reliable.

For comparison, we also presented the WD masses  in DO Dra and XY Ari obtained earlier by other authors without using the X-ray spectroscopy method, which, within the error bars, coincide with both the values obtained by  \citet{Suleimanov.etal:19} and the values obtained by \citet{2021MNRAS.507.5805A, 2024arXiv240717562A}.
The WD mass in XY Ari was obtained from the estimate of its radius based on the egress duration of the X-ray eclipse
\citep{Hellier:97}. \citet{1997ApJ...476..847H} found for DO Dra that the value of the component mass ratio was $q = 0.45\pm 0.05$ and estimated the donor star mass (0.375$\pm 0.014\, M_\odot$) by adopting the empirical zero-age main sequence (ZAMS) mass-radius relation and assuming a Roche lobe filling. \citet{2024arXiv240717562A} obtained different values 
($q = 0.62 \pm 0.02$ and 0.62$\pm 0.07\, M_\odot$).
We note that the obtained donor star mass is almost twice as high
as the value predicted by the standard 
$P_{\rm orb} - M_2$ sequence 
\citep[$\approx 0.32\,M_\odot$ 
at $P_{\rm orb} = 3.97$\,h, ][]{2011ApJS..194...28K}. 

If we believe that the novel systematic differences in the WD masses presented in Table\,\ref{tab:mass}  are real, then an explanation must be found for them. We suggest that this difference is due to the low luminosities of the studied IPs.
Indeed, three out of four  IPs, namely XY Ari, DO Dra, and EX Hya,  have low X-ray luminosities, and therefore low mass accretion rates. Therefore the PSRs above the WD surface could be relatively tall ($H_{\rm sh}/R > 0.1$, $R$ is the WD radius) in these sources. This fact is not favorable for employing the X-ray spectroscopy method to estimate the WD mass because the accretion column height becomes an additional parameter in the model of X-ray spectra of IPs if it is relatively high. Moreover, the contribution of cyclotron cooling becomes more significant for tall columns, leading to additional uncertainties because the magnetic field strength is, as a rule, not known accurately. We have to conclude that the currently used X-ray spectroscopy method can estimate only a lower limit of the WD mass in low X-ray luminous IPs. We note that the WD masses in EX Hya
and DO Dra were obtained by \cite{Suleimanov.etal:19} using the model X-ray spectra grid computed for tall PSRs with a fixed relative height, $H_{\rm sh}/R =0.25$. 
If a standard low PSR grid had been used, the resulting WD masses for these IPs would have been even lower \citep[by approximately 0.07\,$M_\odot$, ][see also Table\,\ref{tab:wd_ip} in this paper]{Suleimanov.etal:19}.

The X-ray spectroscopy method is preferable for the classical high luminous IPs with high accretion rates, i.e., $\sim 10^{-9}\,M_\odot$\, yr$^{-1}$. Indeed, most IPs have high X-ray luminosities \citep{Suleimanov.etal:19, 2022MNRAS.511.4937S}. In such systems the local mass accretion rate is sufficiently high ($a > 10$\,g\,s$^{-1}$\,cm$^{-2}$) and the resulting accretion column height corresponding to the shock position is relatively low, namely, about a few percent of the WD radius. In this case the accretion column height and the cyclotron cooling are negligible. Therefore, we expect that the WD mass determination in high X-ray luminous IPs is more accurate. Unfortunately, it seems that WD masses in the luminous IPs cannot be determined using the method used by \citet{2021MNRAS.507.5805A, 2023arXiv230408524A, 2024arXiv240717562A}, because the optical flux in these systems is dominated by the accretion disk, and the donor star emission can hardly be investigated. Thus, the X-ray spectroscopy method remains the only way for WD mass determination in high luminous IPs, and it is important to study all the possible systematic uncertainties of the method. 

 \begin{table}
\caption{WD masses in IPs obtained from hard X-ray spectral fitting \citep[$M_{\rm x}$,][]{Suleimanov.etal:19} and from non-X-ray spectroscopy methods,
mostly optical ($M_{\rm opt}$).
 \label{tab:mass} 
 }
{\footnotesize
\begin{center} 
\begin{tabular}[c]{ lc| cl  l}
\hline
 Name   &$M_{\rm x}^{\rm a}$& $M_{\rm opt}^{\rm a}$    \\ 
\hline
\hline
 GK Per	&  0.79$\pm 0.01$  &     1.03$_{-0.11}^{+0.16}$$^b$    \\
 XY Ari	&  1.06$\pm 0.10$  &     1.21$\pm 0.04$$^c$    \\
         &                  &     1.04$\pm 0.13^d$ \\
 DO Dra	&  0.76$\pm 0.06$  &     0.99$\pm 0.10^e$ \\
         &                  &     0.83$\pm 0.10$$^f$    \\
 EX Hya	&  0.70$\pm 0.04$  &     0.78$\pm 0.03$$^g$    \\
         &                  &     0.79$\pm$ 0.03$^h$  \\
 \hline
\end{tabular}
\end{center}
}
Notes: a - in solar masses; b - \citet{2021MNRAS.507.5805A}; c - \citet{2023arXiv230408524A};  
d - \citet{Hellier:97};  e - \citet{2024arXiv240717562A}; 
f - \citet{1997ApJ...476..847H}; g - \citet{2016MNRAS.461.1576E}; h - \citet{BR:08}.
\end{table}

In general, a comparison of WD masses obtained by the X-ray spectroscopy method and the optical methods opens the possibility to understand the physics of tall PSRs. In the present paper, we begin by examining external assumptions not connected with the PSR physics which were made in the X-ray spectroscopy method, and we investigate
the potential systematic uncertainties these assumptions may introduce.

The remainder of the paper is organized as follows. In Sect.\,\ref{sect:method} we outline the method for WD mass determination from X-ray spectra. We continue in Sect.\,\ref{sect:uncertainties} by considering the systematic external uncertainties affecting this method. In Sect.\,\ref{sect:grids} we present the calculations of our new model grids of PSR spectra. We present our results in Sect.\,\ref{sect:results} and summarize and conclude our investigations in Sect.\,\ref{sect:discussion}.

\section{Method}
\label{sect:method}
  
The post-shock plasma temperature and, therefore, the hardness of the X-ray spectrum, depends on the plasma velocity before the shock. To first approximation this velocity is a free-fall velocity, $\vv_{\rm ff}$ at the WD surface, and therefore, depends on the WD mass $M$ and radius, $R$,
\be
       kT_{\rm sh} = \frac{3}{16}\mu\,m_{\rm H}\, {\vv_{\rm ff}^2} \sim \frac{M}{R(M)},
\ee  
where $\mu \approx 0.62$ is the mean molecular weight of a fully ionized plasma with solar chemical composition, and $m_{\rm H}$ is the mass of the hydrogen atom. Using the $M-R$ relation suggested by \citet{Nbg:72} for the cold WDs,
\be \label{mrn}
        R = 0.0112 R_\odot \left(\left(\frac{1.44\,M_\odot}{M}\right)^{2/3}-\left(\frac{M}{1.44\,M_\odot}\right)^{2/3}\right)^{1/2},
\ee
we obtain a direct connection between the observed hardness of X-ray spectrum and the WD mass, which can be used for WD mass determination. 
In fact, the plasma in PSRs is optically thin in the X-ray band and it settles down below the shock to the WD surface in the sub-sonic regime, cooling down and becoming denser. Hence, the total PSR X-ray spectrum is softer than the bremsstrahlung spectrum with post-shock temperature $kT_{\rm sh}$   \citep{1995ApJ...455..260W} and a detailed hydrodynamical model of the PSR has to be computed. 

In recent decades, a few one-dimensional PSR models and corresponding X-ray spectra grids were computed and used for WD mass determinations in IPs and polars \citep{1998MNRAS.293..222C, 2000MNRAS.314..403R, SRR:05, Betal:09, Yuasa.etal:10}. In all the cited works the assumptions described above, the free-free velocity before the shock and the Nauenberg's $M-R$ relation as well as  a cylindrical PSR geometry were used.

However, these assumptions could be inaccurate. First of all, the plasma velocity above the shock could differ from the free-fall velocity at the cold WD surface. We point out three main possible reasons:

1) Nauenberg's relation (\ref{mrn}) could be inaccurate, and actual WD radii could be larger with another $R'(M)$;

2) The plasma falls not from the infinity, but from the magnetospheric radius, $R_{\rm m}$, which could be relatively small;

3) The shock could be high enough above the WD surface, at a height, $H_{\rm sh}$.

Therefore, the plasma velocity before the shock, $\vv_0$, could be computed as
\be  \label{eq:all}
       kT_{\rm sh} \sim {\vv_{\rm 0}^2} \sim \frac{M}{R'(M)}\left(\left(1+h_{\rm sh}\right)^{-1}
       -r_{\rm m}^{-1} \right),
\ee
where $h_{\rm sh}=H_{\rm sh}/R$ is the relative shock (and PSR) height, and $r_{\rm m}=R_{\rm m}/R$ is the relative magnetospheric radius.

Another potentially important problem is the dipole geometry of the PSR \citep{2005A&A...440..185C}. Therefore, the PSR cross-section, $S$, increases with the distance from the WD surface, $h$, because the magnetic field lines diverge and the magnetic field strength decreases as $B \sim (R+h)^{-3}$. This effect is especially important for tall PSRs, which can exist at low mass-accretion rates. PSR models in quasi-dipole geometry (assuming that $S(h)\sim (R+h)^{-3}$) were also created and used for IPs' hard X-ray spectra fitting \citep{HI:14a, Suleimanov.etal:16}.

In the latter paper, the possibility that the magnetospheric radius is small was also considered, and a two-parameter model spectra grid was computed, where the second parameter was the relative magnetospheric radius $r_{\rm m}$. However, the magnetospheric radius is not known for most IPs, and we assumed that the magnetospheric radii are equal to the corotation radii to find WD masses in most IPs \citep{Suleimanov.etal:19}. The corotation radius $R_{\rm C}$ is the distance from the WD center, where the Kepler angular velocity is equal to the WD spin angular velocity,
\be
        R_{\rm C} = \left(\frac{GM}{\omega_{\rm s}^2}\right)^{1/3} =\left(GM\frac{P_{\rm s}^2}{4\pi^2}\right)^{1/3}.
\ee    
Here $\omega_{\rm s}$ is the angular rotation velocity of the WD, and $P_{\rm s}$ is the spin period of the WD. In fact, the maximum possible magnetospheric radius in accreting IPs is equal to the corotation radius ($R_{\rm m} = R_{\rm C}$), because it is commonly accepted that accretion is not possible if $R_{\rm m} > R_{\rm C}$. Actual magnetospheric radii can be smaller, as it was found for several IPs with a power law breaking in the power density spectra \citep{Suleimanov.etal:19}. It was assumed that the breaking frequency is equal to the Kepler frequency at the magnetospheric radius \citep{Revnivtsev.etal:09}.

A fraction of the hard X-ray emitting IPs, such as EX Hya and DO Dra, have low luminosities and, therefore, low mass accretion rates. It means that the PSRs in these objects could be tall. A separate grid of IP model spectra with a fixed relative PSR height of $h_{\rm sh} = 0.25$ was computed and used for finding WD masses in low luminous IPs \citep{Suleimanov.etal:19}.

It is thus clear that most sources of potential systematic uncertainties in WD mass determination in IPs, except the $M-R$ relation, were already taken into account. We call them external systematic uncertainties and investigate which corrections we can expect if they are ignored. Here we do not consider uncertainties connected with approximations which are made for computation of the PSR models.
We refer to these uncertainties as internal uncertainties.

\section{External systematic uncertainties}
\label{sect:uncertainties}

Let us consider a simplified approach for estimating the expected corrections to the WD masses. First of all, we generalize all the observational data connected with a given IP hard X-ray spectrum via $(M/R)_{\rm obs}$. The fiducial WD mass for this IP is the mass obtained using Nauenberg's $R(M)$ relation assuming that $r_{\rm m} \rightarrow  \infty$ and $h_{\rm sh} \rightarrow 0$. The corrected WD mass is the mass, which gives the same  $(M/R)_{\rm obs}$ if some finite values of $r_{\rm m}$ and $h_{\rm sh}$ or another $M-R$ relation are used.

\subsection{Improved $M-R$ relation} 
\label{sec:hotmr}

\begin{figure} 
\begin{center}  
\includegraphics[width=  0.9\columnwidth]{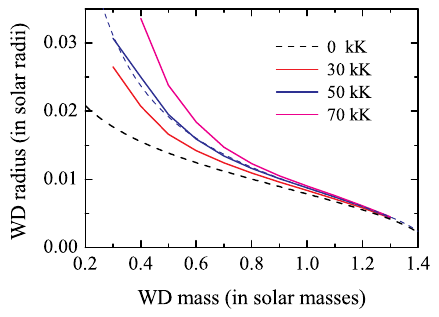}
\caption{\label{fig:mr} 
WD mass-radius relations for different effective temperatures of the hydrogen envelope \citep{2001PASP..113..409F}. The relation suggested by \citet{Nbg:72} for zero temperature WDs is shown with the  dashed black curve. 
The  dashed blue curve is the fit for 50 kK envelope temperature, see Eq.\ref{mrN} and Table\,\ref{tab:mr}.}
\end{center} 
\end{figure}

\begin{figure} 
\begin{center}  
\includegraphics[width=  0.9\columnwidth]{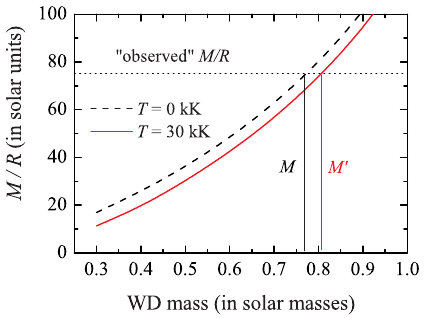}
\caption{\label{fig:mrfix} 
Dependences of the $M/R$ ratio on the WD mass for different envelope temperatures, $T=0$ (dashed black curve), and $T=30$\,kK (solid red  curve). For  fixed $M/R$, the second relation gives a larger WD mass.
}
\end{center} 
\end{figure}

The validity of Eq.\,(\ref{mrn}) can be tested, for example, by observing isolated WDs. Independent determinations of masses and radii of such WDs are performed using different approaches \citep[see, e.g.,][and references therein]{2018MNRAS.479.1612J, 2019MNRAS.484.2711R}. As a rule, the measured WD radii are larger than the radii evaluated from Eq.\,(\ref{mrn}) \citep[see, e.g.,][and papers cited above]{2017MNRAS.470.4473P}, especially for low-mass WDs. The reason is the finite temperature of WDs and the presence of hydrogen or helium envelopes with finite temperatures. Indeed, computations of WD radii with finite WD temperatures and relatively thick hydrogen envelopes with a constant relative hydrogen envelope mass $M_{\rm H}/M_{\rm WD} = 10^{-4}$ showed a similar WD radii increase \citep{1995LNP...443...41W, 2001PASP..113..409F}.    
 
In fact, the computed WD radius depends on the envelope chemical composition and its relative mass and, apparently, the envelopes of high-mass single WDs ($\gtrsim 0.8\,M_\odot$)  are thin, $\sim 10^{-6}\,M_\odot$ \citep[see, e.g.,][]{2019MNRAS.484.2711R}. 
But it is obvious that the thickness and chemical composition of the envelopes for accreting WDs in CVs differ from the parameters of the envelopes of isolated WDs. Due to the constant influx of matter with approximately solar chemical composition, episodes of explosive thermonuclear burning in the accreted envelope periodically occur on the WD surface, which we observe as nova outbursts.
The mass of the accumulated envelope required to initiate a thermonuclear outburst is between $10^{-7}$ 
and $10^{-3}\,M_\odot$, and depends on the WD mass and the mass accretion rate
\citep[see recent review][for details]{2021ARA&A..59..391C}. This critical envelope mass is about 
$10^{-4}\,M_\odot$ for a typical high luminous IP with WD mass $ \sim 0.8\,M_\odot$ and a mass accretion rate $\sim 10^{-9}\,M_\odot$\,yr$^{-1}$.

Most of the envelope is carried away during the outburst, but some fraction remains, because after the envelope becomes transparent, the post-novae are observed to have a super-soft source X-ray phase 
\citep[see e.g.,][]{2001A&A...373..542O}. Thermonuclear burning continues in the remaining helium-rich envelope \citep[see e.g.,][]{Rauch_etal2010, 2024A&A...689A.335T}.
The thermonuclear burning during the super-soft phase necessarily leaves a layer of helium-rich ash that accumulates over time \citep[see e.g.][]{2013ApJ...762....8D}.

Thus, at present we can only speculate about the thickness and chemical composition of the accreted envelopes on the WD surfaces in IPs. The correct answer for each specific IP will depend on what 
fraction of the envelope ultimately  remains on the WD surface after the nova outburst and on the time that has passed since the previous outburst. Therefore, we decided to consider thick hydrogen envelopes as a first approximation and to investigate what the effect this would have on the determination of WD masses by the X-ray spectroscopy method. For this aim we use 
WD models with thick hydrogen envelope, computed on the base of the \citet{2001PASP..113..409F} models and presented on the website of the Montreal White Dwarf Database (MWDD)\footnote{\url{https://www.montrealwhitedwarfdatabase.org/evolution.html}}. Examples of $M-R$ relations computed for various WD effective temperatures are presented in Fig.\,\ref{fig:mr}. 

\begin{figure} 
\begin{center}  
\includegraphics[width=  0.9\columnwidth]{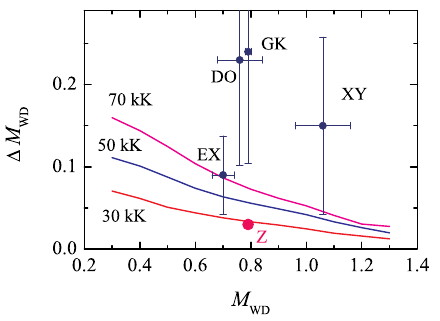}
\includegraphics[width=  0.85\columnwidth]{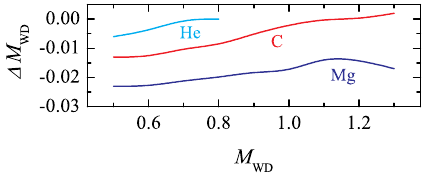}
\caption{\label{fig:dmt} 
{\it Top panel:} Corrections to the WD masses determined using zero temperature WD mass-radius relation \citep{Nbg:72} depending on the hydrogen envelope temperature. Differences between WD masses  found using hard X-ray observations \citep{Suleimanov.etal:19} and using optical methods for four objects are also shown. The difference between the average WD mass in IPs and all the CVs \citep{Zorotovic.etal:11} is shown with the pink dot. {\it Bottom panel:} Corrections to the WD masses determined using the \citep{Nbg:72}  mass-radius relation (\ref{mrn}) and numerical M-R relations obtained by \citet{HS61} for different WD chemical compositions.
 }
\end{center} 
\end{figure}

Let us now estimate what error we introduce when using Eq.(\ref{mrn}) instead of the relations shown in Fig.\,\ref{fig:mr}.  Our estimations are based on the assumption
that the emergent spectrum of the PSR depends only on the ratio $M/R$, because $\vv_0^2 \sim M/R$. This means that we assume that any observed hard X-ray spectrum of some IP corresponds to
some ratio $M/R$. Let us call it the ``observed'' $M/R$ ratio. 
In order to find the WD mass  from the pseudo-observed $M/R$ ratio we must use some model relation between the mass and radius of the WD. Previously, the relationship given by Nauenberg's formula 
(Eq.\,\ref{mrn}) was always used. It is clear that if we use another relation calculated for the finite WD envelope temperature, we will get a WD mass higher than using Eq.\,\ref{mrn}, because $R(T)> R(T=0)$; see Fig.\,\ref{fig:mrfix}. 

Then, we compute, for each WD mass within the considered range ($0.3-1.3$ solar masses) the ratio $M/R(M,T=0)$, where the zero-temperature radii $R(M,T=0)$ are obtained from Eq.\,(\ref{mrn}). In the next step, we fix
the ratio $M/R(M,T=0)$ as an ``observed'' $M/R$ ratio. We then determine at which WD mass this ``observed'' ratio matches the ratio $M/R(M,T)$ computed using WD radii at a finite envelope temperature (see Fig.\,\ref{fig:mrfix}). This new WD mass $M'$ is higher than the initial one $M$, and the difference  between these masses $\Delta M = M'-M$ is the error we are looking for. The dependencies of the estimated errors $\Delta M$ for different $R(M,T)$ are shown in Fig.\,\ref{fig:dmt}. The differences between WD masses obtained using hard X-ray spectra \citep{Suleimanov.etal:19} and using the optical methods (see Table~\ref{tab:mass}) are also shown for four IPs in the same figure.

Fig.\,\ref{fig:dmt} shows that the difference between optical and X-ray WD mass measurements for GK\,Per, XY\,Ari, DO\,Dra, and EX\,Hya can be partially explained with sufficiently high surface temperatures of the WDs in these systems. An especially high WD surface temperature, above 50\,kK, is expected for GK\,Per. This system was Nova Persei 1901 and we can assume that the WD surface is still hot after the nova explosion. This IP is also a dwarf nova and the magnetospheric radius in the system could be small during outbursts \citep{Suleimanov.etal:16, Suleimanov.etal:19, Wada.etal:18}. We devote a separate paper to this IP because of its great importance. We conclude that a finite WD surface temperature leads to an underestimation of WD masses if Eq.(\ref{mrn}) is used, and can introduce  systematic error when determining WD masses in IPs using hard X-ray spectroscopy.

The actual WD temperatures in CVs were investigated by \citet{2009ApJ...693.1007T}. They found that the WDs are heated due to accretion, and their temperatures vary from 10 to 50\,kK. The actual value for a given CV depends on the average mass accretion rate $\langle\dot M\rangle$. They derived an equation which describes this dependence:
\be \label{eq:taccr}
      T_{\dot M} \approx 19\,{\rm kK}\,\langle\dot M_{-10}\rangle^{1/4}\,m,
\ee
where $\dot M_{-10} = \dot M/ 10^{-10}$ M$_\odot$\,yr$^{-1}$, and $m=M/M_\odot$. The mass accretion rates in most of the investigated IPs are close to $10^{-9} M_\odot$\,yr$^{-1}$ \citep{Suleimanov.etal:19}. Therefore, we expect that the WD temperatures in IPs are close to 30\,kK.

We note that Nauenberg's relation (\ref{mrn}) is also approximation of the numerical computations of the cold WD models with various chemical compositions of WDs
performed by \citet{HS61}. The direct comparison of the numerical results presented in Table 1A by \citet{HS61} and Eq.\,\ref{mrn} shows that the approximation formula very accurately describes the carbon WD models at masses above one solar masses, and the helium WD models at lower masses. The approximation formula overestimates the radii of carbon WD at low ($< 1\,M_\odot$) masses, and the radii of magnesium WD at all masses. This leads to the fact that the WD masses are slightly overestimated (when Eq.\ref{mrn} used) by $0.01 - 0.015\,M_\odot$ at masses below one solar masses compared to using numerical models of carbon WDs and by about $0.02\,M_\odot$ for magnesium WDs (see Fig.\,\ref{fig:dmt}, bottom panel).

\subsection{Importance of WD rotation} 

The accretion flow moving along the magnetic field lines of a rotating magnetized WD is affected by the centrifugal force, because the magnetosphere rotates with the same angular velocity as the WD. The influence of the WD rotation can additionally decrease the plasma velocity at the WD surface.

First, let us consider a rotating magnetic monopole, assuming that the magnetic field lines are radial and straight, and the rotation axis is normal to the orbital plane. In this case the momentum conservation equation can be written as
\be
      {\vv} \frac{d\vv}{dr} = -\frac{GM}{r^2}+\omega_{\rm s}^2\,r.
\ee
The solution of this equation is
\be
      {\vv_0^2}  = \frac{2GM}{R}\left(1-r_{\rm m}^{-1}\right)-\omega_{\rm s}^2\,R^2\left(r_{\rm m}^2-1\right),
\ee
or in another form
\be \label{eq:rot}
      {\vv_0^2}  = {\vv_{\rm ff}^2}\left(1-r_{\rm m}^{-1}-\frac{1}{2}r_{\rm C}^{-3}\left(r_{\rm m}^2-1\right)\right),
\ee
where $r_{\rm C} = R_{\rm C}/R$. It is clear that the assumption $R_{\rm m}=R_{\rm C}$ leads to 
\be
       {\vv_0^2}  = {\vv_{\rm ff}^2}\left(1-\frac{3}{2}r_{\rm m}^{-1}\right)
\ee
instead of ${\vv_0^2}  \sim \left(1-r_{\rm m}^{-1}\right)$ used before (see Eq.\,(\ref{eq:all})).

\begin{figure} 
\begin{center}  
\includegraphics[width=  0.9\columnwidth]{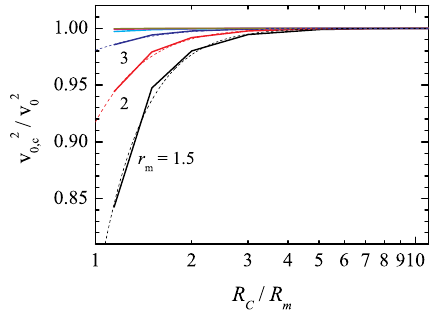}
\caption{\label{fig:relV} 
Dependence of the square of the ratio of the plasma velocities at the WD surface, computed by integration of Eq.\,(\ref{eq:dip1}) and predicted by Eq.\,(\ref{eq:rot}), on the ratio $R_{\rm C}/R_{\rm m}$ for various values of $r_{\rm m}$. Approximations obtained using Eq.\,(\ref{eq:fit}) are also shown with dashed curves. }
\end{center} 
\end{figure}

\begin{figure} 
\begin{center}  
\includegraphics[width=  0.8\columnwidth]{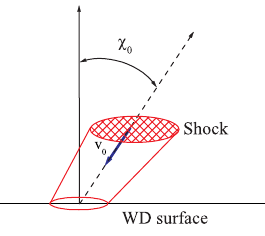}
\caption{\label{fig:oblq} 
Geometry of the oblique accretion flow at the WD surface.}
\end{center} 
\end{figure}

\begin{figure} 
\begin{center}  
\includegraphics[width=  0.9\columnwidth]{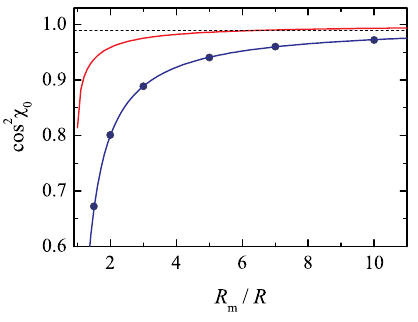}
\caption{\label{fig:incl} 
Dependence of the magnetic field line inclination at the WD surface in the case 1 (coaxial rotator, blue curve, Eq.\,\ref{eq:incl1}) and the average inclination angle computed using Eq.\,(\ref{eq:incl2}) (red curve). The blue dots are the results of the numerical integration for case 1.
 }
\end{center} 
\end{figure}

However, the real magnetic field is rather a dipole, and the influence of the magnetic field line curvature can be important. The dynamics of an accretion flow in the dipole geometry is presented in Appendix\,\ref{app:dip}. By integrating Eq.\,(\ref{eq:dip1}) we found that the Eq.\,(\ref{eq:rot}) is also accurate enough except for the cases with $r_{\rm m} < 3$ and, simultaneously, $r_{\rm C} < 3 r_{\rm m}$ (see Fig.\,\ref{fig:relV}). In these cases the computed velocity at the WD surface $\vv_{\rm 0,c}$  is smaller than that predicted by Eq.\,(\ref{eq:rot}), $\vv_{0}$,
\be \label{eq:RF}
        {\vv_{\rm 0,c}^2} = RF \,\vv_{0}^2, 
\ee
where the reduction factor, $RF$, can be approximated as
\be \label{eq:fit}
     RF = 1-\frac{1}{\left(r_{\rm  Cm}+a\right)^b},
\ee 
where $r_{\rm  Cm}= R_{\rm C}/R_{\rm m}$, $a=-0.38+0.53\,r_{\rm m}$, and $b=3.7+0.44\,r_{\rm m}$.

\subsection{Importance of accretion flux inclination}
\label{sect:incl}

The inclination, $\chi$, of the accretion flow at the WD surface ($\chi=\chi_0$) can be potentially important (see Fig.\,\ref{fig:oblq}). Without a magnetic field, only the normal component of the plasma velocity participates in the formation of the shock wave, while the tangential component will remain constant. Therefore, the plasma flow must change direction of the movement \citep{LL6}. However, in our case, the plasma is frozen to the magnetic field and cannot change direction of the movement. As a result, we assume that the plasma passing through an oblique shock wave is heated and changes its velocity and density in the same way as when passing a normal shock wave. As a result, we only make corrections to the increase in the footprint area of the PSR ($S'_0 = S_0/\cos\chi_0$) and to the decreasing of the gravity, $g'(z) = g(z) \cos \chi_0$ 
\citep[see equations 22 and 23 in][]{Suleimanov.etal:16}.

The value of $\chi_0$ depends on the angle between the rotation axis and the magnetic axis. In Appendix\,\ref{app:dip} we consider two limiting cases, namely, that both axes coincide (a coaxial rotator, case 1), and that the magnetic axis is normal to the rotation axis (an orthogonal rotator, case 2). For case 1 the value sought is
\be  \label{eq:incl1}
      \cos^2 \chi_0 = \frac{1-\cos^2 \lambda_0}{1-0.75\,\cos^2 \lambda_0} =   \frac{1-r_{\rm m}^{-1}}{1-0.75\,r_{\rm m}^{-1}} ,
\ee 
where $\lambda_0$ is $\lambda$ at the WD surface, $\cos^2 \lambda_0 = R/R_{\rm m}=r_{\rm m}^{-1}$ (see Appendix\,\ref{app:dip}). The same expression is correct for the magnetic field line which is tangent to the magnetospheric radius in case 2.

In fact, we do not know the angle between magnetic axis and the rotation axis. Hence, we assume that the average value of $\chi_0'$ corresponds to the angle $\lambda_0' = 0.5(\lambda_0+\pi/2)$, which corresponds to the middle point between angles $\lambda_0$ and $\pi/2$. In this case the required value is
\be \label{eq:incl2}
      \cos^2 \chi_0' =   \frac{1+(1-r_{\rm m}^{-1})^{1/2} }{1.25+0.75(1-\,r_{\rm m}^{-1})^{1/2} }.
\ee 
The value of $\cos^2 \chi_0'$ is less than 0.99 for $r_{\rm m} < 6$ only (see Fig.\,\ref{fig:incl}). Therefore, we have to take into account the magnetic field inclination for very small magnetospheres only. These cases have to be considered separately, and for the determination of WD masses in relatively slow rotating IPs (with $P_{\rm s} > 200$\,s) it is sufficient to use Eq.\,(\ref{eq:rot}).

\begin{figure} 
\begin{center}  
\includegraphics[width=  0.9\columnwidth]{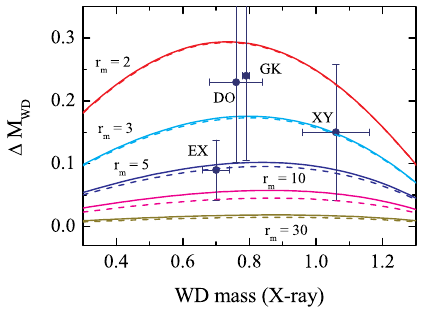}
\caption{\label{fig:dmrm} 
Corrections for the WD masses determined using zero temperature WD mass-radius relation \citep{Nbg:72} depending on the relative magnetospheric radius, $r_{\rm m}$, with $r_{\rm C} = 10$ or $r_{\rm C} = 1.25 r_{\rm m}$ if $r_{\rm m} \ge 10$ (solid curves). Corrections computed for non-rotating WDs are shown with dashed curves. Differences between WD masses found using hard X-ray observations \citep{Suleimanov.etal:19} and using optical methods are displayed for four IPs. 
 }
\end{center} 
\end{figure}

\subsection{Importance of magnetospheric radius}

We estimated the errors in WD mass determination if the finite magnetospheric radius and the WD rotation are ignored. For this aim we used a correction to $\vv_0^2$ presented in Eq.\,(\ref{eq:rot}). The correction was computed for WD masses from $0.3-1.3$\,$M_\odot$, and a few values of the relative magnetospheric radius $r_{\rm m}$ (see Fig.\,\ref{fig:dmrm}). The value of $r_{\rm C}$ was fixed to be 10, if $r_{\rm m} < 10$. Otherwise the value was taken to be equal 1.25\,$r_{\rm m}$. The correction was found using the following method. First of all we computed the relation $M - M/R(M)$ using Eq.\,(\ref{mrn}). Then for every $M$ we corrected the corresponding value $M/R(M)$ by multiplying with the ratio $\vv_{\rm ff}^2/\vv_0^2$, where $\vv_0^2$ was computed using Eq.\,(\ref{eq:rot}). Then using this corrected ratio $(M/R(M))'$ we found a new value of the WD mass $M'$ from the relation $M - M/R(M)$. The final correction is $\Delta M = M'-M$.

The correction is significant for small $r_{\rm m}$, but the importance of WD rotation at the chosen $r_{\rm C}$ is not. A finite magnetospheric radius was already taken into account in the IP spectra models \citep{Suleimanov.etal:16}. However, the direct $R_{\rm m}$ estimations from the breaking frequency in the power spectra are possible for a few IPs only, and the assumption $R_{\rm m} = R_{\rm C}$ was used for determination of WD masses \citep{Suleimanov.etal:19, Shaw.etal:20}. This assumption also leads to WD mass underestimations, because magnetospheric radii can be smaller than co-rotational radii. This underestimation must be significant if $r_{\rm m}$ is small and we do not have any information about it. In particular, the difference between $M_{\rm opt}$ and $M_{\rm x}$ for XY\,Ari can be partially explained by a small magnetospheric radius. The relative corotation radius in this system is about 10 \citep{Suleimanov.etal:19}, and the underestimation of the mass can be as high as 0.1\,M$_\odot$, if the actual value of $r_{\rm m}$ is less than five.   The influence of WD rotation and inclination of the accretion columns relative to the normal to the WD surfaces were not taken into account before.

\begin{figure} 
\begin{center}  
\includegraphics[width=  0.9\columnwidth]{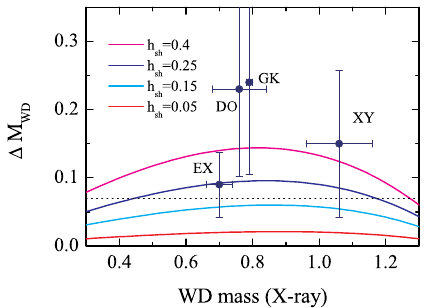}
\caption{\label{fig:dmh} 
Corrections to the WD masses determined using zero temperature WD mass-radius relation \citep{Nbg:72} depending on the relative accretion column height $h_{\rm sh}$. Differences between WD masses  found using hard X-ray observations \citep{Suleimanov.etal:19} and using optical methods  are shown for four IPs. The mass correction (0.07\,M$_\odot$) obtained using tall PSR models \citep[$h_{\rm sh}=0.25$, ][]{Suleimanov.etal:19} is shown with the  dotted black line.
 }
\end{center} 
\end{figure}

\subsection{Importance of accretion column height} 

As a first approximation, the shock height above the WD surface is inversely proportional to the local mass accretion rate $a= \dot M / 2S_{\rm PSR}$
\citep{Wu.etal:94}, where $\dot M$ is the total mass accretion rate, and $S_{\rm PSR}$ is the area of the accretion column footprint. The accretion column height becomes significant at $h_{\rm sh} \gtrsim 0.05$, i.e., at $a \lesssim 10$\,g\,s$^{-1}$\,cm$^{-2}$, and this fact can be important for WD mass determination. Unfortunately, the PSR footprint areas are poorly known, and the values of $a$ have large uncertainties. We can only assume that $a$ is smaller for low luminous IPs. This is the reason why we used a separate grid computed with a fixed $h_{\rm sh} = 0.25$ to determine the WD masses in low luminous IPs, such as EX\,Hya and DO\,Dra \citep{Suleimanov.etal:19}. 
 
We estimated the errors in the determination of the mass of WD at various values, $h_{\rm sh}$ (see Fig.\,\ref{fig:dmh}). The same method as was described in the previous subsection was used for this aim. The only difference is that $\vv_0^2$ was computed from ${\vv_{\rm ff}^2}(1+h_{\rm sh})^{-1}$ instead of using Eq.\,(\ref{eq:rot}). The corrections due to finite accretion column heights are less significant than the corrections due to  finite magnetospheric radii. However, we note that the mean difference between the WD masses obtained using the short and tall ($h_{\rm sh} = 0.25$) column models is about 0.07\,M$_\odot$ \citep[see fig.11 in ][]{Suleimanov.etal:19}. Hence, the simple estimations presented in Fig.\,\ref{fig:dmh} overestimate the effect of column height due to the influence of divergent magnetic field lines for tall columns.
 
The masses of WD in EX\,Hya, and DO\,Dra were determined using the model grid with $h_{\rm sh} = 0.25$. This means that the differences $M_{\rm opt}-M_{\rm x}$ for these objects probably arise due to the finite surface temperatures of WD. On the other hand, a new estimation of the distance to XY\,Ari 
\citep[308$\pm$40\,pc,][]{2023arXiv230408524A} is considerably smaller than the assumption that the distance is 2\,kpc, which was made by \citet{Suleimanov.etal:19}. Hence, the bolometric luminosity of XY\,Ari is less than we estimated before, $L \approx 7\times 10^{32}$\,erg\,s$^{-1}$, and is comparable with the luminosity of DO\,Dra. Therefore, the height of the accretion column in this IP could be significant and could affect the determination of the WD mass. 

\section{The new grid}
\label{sect:grids}

We computed a new grid of PSR spectra assuming a finite temperature of the WDs, see below. The current models are based on the models described in \citet{Suleimanov.etal:16}. We used the same range of WD mass, from 0.3 to 1.4\,M$_\odot$ with a step of 0.02\,M$_\odot$, and the same values for the relative magnetospheric radius $r_{\rm m}$ from 1.5 to 60 calculated as $60/N$, where $N$ changes from 40 ($r_{\rm m} = 1.5$) to 1 ($r_{\rm m} = 60$); a total of 40 sets. However, two additional sets with $r_{\rm m} = 100$ and 1000 were not included in the grids, because they are practically not different from the models with $r_{\rm m} = 60$. We still used fixed values of the relative WD area occupied by the PSR footprint, $2S_{\rm PSR} / 4\pi R^2= 5 \times 10^{-4}$, and a constant mass accretion rate, $\dot M = 10^{17}$\,g\,s$^{-1}$.

Based on the results presented in the previous section, we introduced the following changes to the model:

1) We used a different $M-R$ relation computed for a thick hydrogen envelope with temperature 30\,kK.
The numerical relation was approximated with the modified Nauenberg formula
\be \label{mrN}
        R = A\, R_\odot \left(\left(\frac{1.44\,M_\odot}{M}\right)^{\gamma}-\left(\frac{M}{1.44\,M_\odot}\right)^{\gamma}\right)^{1/2},
\ee
where the parameters $A$ and $\gamma$ depend on the effective temperature of the hydrogen envelope and are presented in Table~\ref{tab:mr}. However, this approximation is not accurate enough for $M > 0.5 M_\odot$, and we performed a direct interpolation employing numerical $M-R$ relation computed using the MWDD.

 \begin{table}
\caption{Dependence of the parameters of modified Nauenberg relation on the effective temperature of the hydrogen envelope. 
 \label{tab:mr} 
 }
{\footnotesize
\begin{center} 
\begin{tabular}[c]{ c c c c c c c c}
\hline
 T\,(kK)  &  10 & 20 & 30 & 40 & 50 & 60 & 70 \\ 
\hline
\hline
 A\,($10^{-2}$) &  1.01 & 0.85& 0.74& 0.68& 0.63& 0.60& 0.56   \\
 $\gamma$ & 0.84 & 1.22& 1.59& 1.89& 2.17& 2.48& 2.78   \\
  \hline
\end{tabular}
\end{center}
}
\end{table}

\begin{figure} 
\begin{center}  
\includegraphics[width=  0.9\columnwidth]{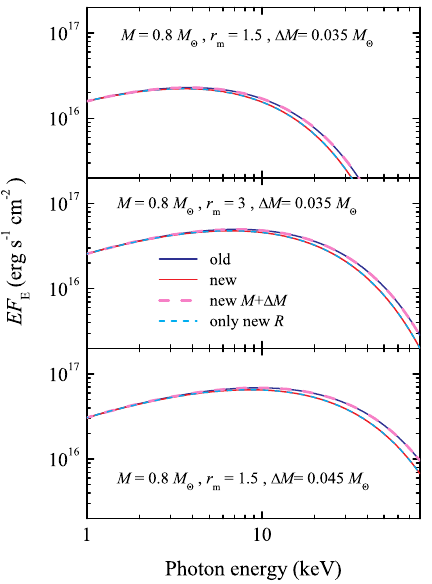}
\caption{\label{fig:sp} 
Comparison of the model spectra computed using the old \citep[blue curves,][]{Suleimanov.etal:16} and the new (red curves) methods for $M$=0.8\,M$_\odot$ and $r_{\rm m}$\,=\,1.5 (top panel), 3 (middle panel), and  10 (bottom panel). The spectra computed with the new method with the increased $M$, which simulate the spectra of the old models, are also shown with the dashed magenta curves (almost coincide with the blue curves). The spectra of the models with new $R$ only are shown with the dashed cyan curves  (almost coincide with the red curves). All model spectra are normalized to the old models' flux at 1\,keV.
 } 
\end{center} 
\end{figure}

\begin{figure} 
\begin{center}  
\includegraphics[width=  0.9\columnwidth]{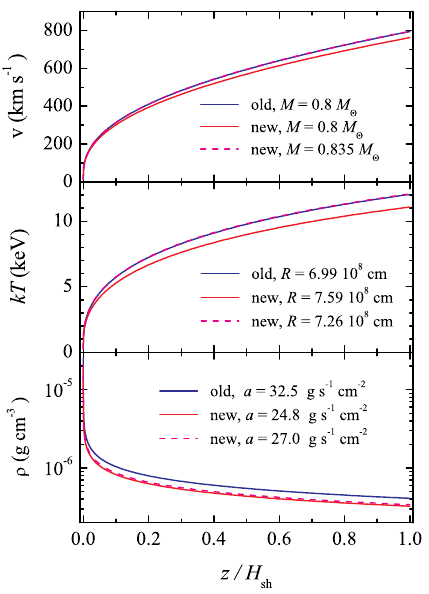}
\caption{\label{fig:mod} 
Comparison of velocity, $\vv$, temperature, $T$, and plasma density, $\rho$ in the PSR models computed using the old \citep[blue curves,][]{Suleimanov.etal:16} and the new (red curves) methods for $M$=0.8\,M$_\odot$ and $r_{\rm m}$\,=\,1.5. The distributions for the new model with the increased $M=0.835$\,M$_\odot$, whose spectrum simulates the spectrum of the old model, are also shown with dashed magenta curves. Parameters of the models are shown in the panels. The shock heights $H_{\rm sh}$ are 4.51$\times 10^5$\,cm, 3.94$\times 10^5$\,cm, and 4.77$\times 10^5$\,cm for the new, old, and new ($M=0.835\,M_{\odot}$) models, respectively.
}
\end{center} 
\end{figure}

2) We used a higher mass accretion rate, $\dot M = 10^{17}$\,g\,s$^{-1}$, instead of $10^{16}$\,g\,s$^{-1}$. The old value gave a significant PSR height for massive WDs, up to $h_{\rm sh} = 0.2$. The adopted higher mass accretion rate allows to keep one a low PSR height even for high-mass WDs.

3) We took into account WD rotation in IPs and used Eq.\,(\ref{eq:rot}) for computing the plasma velocity above the shock with a finite value of the relative corotation radius $r_{\rm C}$. It was fixed to the maximum value among $r_{\rm C} = 10$ and $r_{\rm C} = 1.25\, r_{\rm m}$.
Formally, we also took into account a correction presented by Eq.\,(\ref{eq:fit}), although it is always $>0.99$ at the considered conditions. We also took only a half of the correction presented by the second term on the right-hand side of the equation. This is because the correction was computed for the coaxial rotator case and a dipole magnetic field, but the correction has to be smaller for the other angles. 
The final expression for the velocity of the matter just before the shock is
\be
   {\vv}^2 = RF\frac{2GM}{R'(M)}\left( \left(1+h_{\rm sh}\right)^{-1}-r_{\rm m}^{-1}
   -\frac{1}{2} r_{\rm C}^{-3}\left(r_{\rm m}^2-\left(1+h_{\rm sh}\right)^2\right)\right).
\ee
Despite the fact that the relative heights of the PSR in the models $h_{\rm sh}$ are small, they were still taken into account.

4) We took into account the inclination of the magnetic field lines relative to the WD surface normal. This angle depends on the angle between the rotation axis and the magnetic dipole axis, but the latter one is, as a rule, not known. Hence we took a simple average between the squares of the cosines computed for two limiting cases, namely, the magnetic axis is parallel to the rotation axis (Eq.\,\ref{eq:incl1}) and the case of an orthogonal rotator 
(Eq.\,\ref{eq:incl2}),
\be \label{eq:incl3}
   \cos^2 \bar \chi_0 =(\cos^2 \chi_0+\cos^2 \chi_0')/2 \mbox{ .}
\ee
Therefore, we reduced the local mass accretion rate
as $a'=a \cos \bar \chi_0$, because of the footprint area increase, see Sect.\,\ref{sect:incl}.
We also decreased the gravity by multiplying with the same cosine \citep[see equations 22 and 23 in][]{Suleimanov.etal:16}.

We present some examples of the model spectra computed using the new approach for the hydrogen envelope temperature 30\,kK and compare them with the old model spectra (Fig.\,\ref{fig:sp}). It is clear that we have to 
slightly increase the WD mass for obtaining the same spectrum. Obviously, the effect of the finite WD temperature (and, therefore, the new $M - R$ dependence) is most important. Most of the IPs investigated before have a relatively large corotation radius \citep[and, therefore, relative magnetospheric radii, see][]{Suleimanov.etal:19}, and we expect that the finite WD temperatures will give the main correction to the WD mass determinations. The influence of the WD rotation is less significant because of a relatively large corotation radius.
 
The difference between physical parameters of the PSRs for the models with the smallest $r_{\rm m} = 1.5$ are shown in Fig.\,\ref{fig:mod}. The values of the used WD radii and the local mass accretion rates are also shown. The new model with increased WD mass, which mimics the old model, has almost identical temperature and velocity distributions in comparison with the old model, but the plasma density in the PSR is a little bit higher. 

\section{Results}
\label{sect:results}

To illustrate the importance of the suggested external improvements to the PSR spectral models, we use the large sample of IPs observed by \textit{Swift}/BAT which is presented in the 105-month BAT catalog \citep{oh18}. The hard X-ray spectra of these 47 IPs were fit using the old {\sc ipolar} grid, available in XSPEC, and the new one, computed using the new $M-R$ relation for WDs with the finite envelope surface temperature of 30\,kK. Besides the new $M-R$ dependence, we assumed that the relative magnetospheric radius is smaller than the relative corotation radius by one quarter ($r_{\rm m}=0.75\,r_{\rm C}$). The corrections for the centrifugal force (Eqs.\,\ref{eq:rot} and \ref{eq:fit}) and the accretion stream inclination relative to the WD surface (Eq.\,\ref{eq:incl3}) were also taken into consideration as described in the previous section.
This new grid {\sc ipolar30kk} is also available in XSPEC\footnote{https://heasarc.gsfc.nasa.gov/docs/xanadu/xspec/models/ipolar30kk.html}.

\begin{figure} 
\begin{center}  
\includegraphics[width=  0.9\columnwidth]{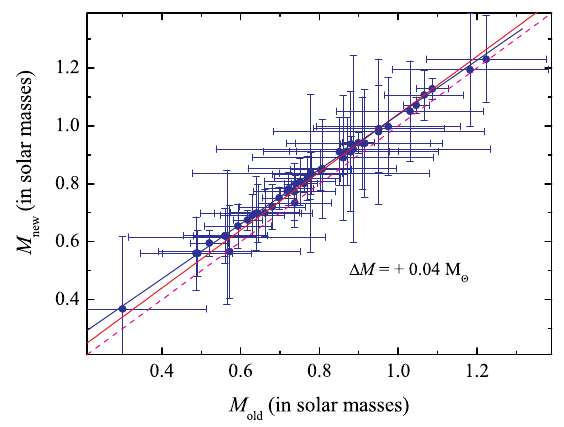}
\caption{\label{fig:mwdc} 
Comparison of the WD masses from the {\it Swift}/BAT sample found using the old published spectral model grid {\sc ipolar} and the new model grid. Lines of equal masses (dashed red) and shifted to 0.04\,$M_\odot$ (solid red) are also shown. A linear fit is displayed by the blue line.}
\end{center} 
\end{figure}

\begin{figure} 
\begin{center}  
\includegraphics[width=  0.9\columnwidth]{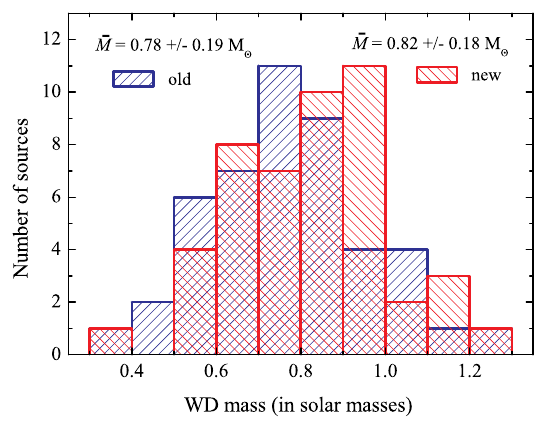}
\caption{\label{fig:mdistr} 
WD mass distributions for the {\it Swift}/BAT sample obtained with the old ({\sc ipolar}) and the new model grids.
  }
\end{center} 
\end{figure}

The investigated IPs and the results of their WD masses determination are presented in Table~\ref{tab:wd_ip}. The observed fluxes in the energy band 14--195\,keV, the distances to the sources, and the derived IP luminosities in the same energy band are also presented. In addition, the known WD spin periods are indicated and the derived relative magnetospheric radii ($r_{\rm m}=0.75\,r_{\rm C}$) are presented. In fact, the WD mass and $r_{\rm m}$ are found simultaneously. Since $r_m$ is the relative radius of the magnetosphere, 
$r_m=R_m/R$, a change in the mass of the WD leads to a change in both its radius, $R$, and the corotation radius ,$R_C$ (note that we accepted $R_m=0.75 R_C$), and therefore $r_m$.

Comparison of the WD masses obtained by using the old and the new grids are shown in Fig.\,\ref{fig:mwdc}. It is obvious that the new WD masses are slightly higher (by approximately 0.04\,$M_\odot$) than the old values, and this difference is larger for low-mass WDs. The value of the difference and the dependence on the WD mass indicate that the main reason for the differences is the new $M-R$ dependence, which takes into account the hydrogen WD envelope with the finite surface temperature of 30\,kK (see Fig.\,\ref{fig:dmt}). Indeed, the used values of $r_{\rm m}$ are close to 10 or larger, see Table~\ref{tab:wd_ip}. Therefore, we expect that the new $M-R$ relation will give the main correction to WD masses.
The obtained WD mass distributions using the both spectral model grids are presented in Fig.\,\ref{fig:mdistr}.    

The average WD mass in the sample is different when using WD masses obtained with the old or the new model grids. Namely, the old average WD mass 0.78$\pm$0.19\,$M_\odot$ is close to the average WD mass in IPs obtained by \citet{Suleimanov.etal:19} (0.79$\pm 0.16\,M_\odot$) and \citet{Shaw.etal:20} (0.77$\pm 0.12\,M_\odot$). The average WD mass obtained with the new grid is 0.04\,$M_\odot$ higher, 0.82$\pm 0.18\,M_\odot$, and it is close to the average WD mass in CVs obtained by the optical methods, 0.82$\pm0.15\,M_\odot$ \citep{Zorotovic.etal:11} and 0.81$^{+0.16}_{-0.20}\,M_\odot$, \citep{2022MNRAS.510.6110P}. Therefore, the previous minor systematic shift between the average WD mass in IPs obtained using the X-ray spectroscopy method and that obtained by the optical methods could be explained by the finite temperature of the WD envelopes.  

We conclude that there are no noticeable systematic differences between the mean WD mass in IPs obtained by X-ray spectroscopy and the mean WD mass in CVs obtained by optical methods.
This means that our assumptions that the accretion columns in the IPs have a negligibly small height and the radii of the magnetospheres are close to the corotation radii are, on average, correct.
However, among the studied IPs there are two groups of objects for which the assumptions made during modeling may turn out to be incorrect, and the estimates of the WD masses obtained for them must be considered as lower limits.

There are three CVs {in our sample, which are, most probably, IPs, but their spin periods are not yet known: IGR\,J12123-5802 \citep{Bernardini.etal:13}, IGR\,J16167-4957 \citep{2011A&A...526A..77B}, and V2487 Oph \citep{2002Sci...298..393H}. We assumed that the value of $r_{\rm m}$ in these systems is equal to the maximum value in the grid, namely 60. In reality $r_{\rm m}$ in these IPs could be less, about 10, as in other high luminous systems (see Table~\ref{tab:wd_ip}). Therefore, the obtained WD masses in these IPs are lower limits as well, and could be approximately 0.05\,$M_\odot$ higher (see Fig.\,\ref{fig:dmrm}). They are also presented as a separate group in Table~\ref{tab:wd_ip}. 

The most interesting source among these three objects is V2487 Oph. This system hosts the most massive WD in our sample (1.23$\pm 0.15 M_\odot$). It is a recurrent nova discovered in 1998 during outburst and has a recurrent time of about 98 years or even shorter \citep{Pagnotta_2009}. The WD mass in the system was estimated as high as 1.35\,$M_\odot$ by \citet{2002ASPC..261..629H} based on the outburst light curve analysis. Formally, this value is comparable with our estimation within errors. However, 
\citet{2024ApJ...969...34S} found that the maximum possible accretion disk truncation in the system is 0.04\,$R_\odot$ ($r_{\rm m} < 7.6$ at $M=1.23\,M_\odot$). Therefore, the correct WD mass could be higher than  $1.23\,M_\odot$.
 
Most of the IPs have high luminosities in the hard X-ray band, exceeding $10^{33}$\,erg\,s$^{-1}$
\citep[see Table\,\ref{tab:wd_ip} and][]{2022MNRAS.511.4937S}. But there are several IPs whose luminosity is lower than this value, and even lower than $10^{32}$\,erg\,s$^{-1}$. They are highlighted in a separate group 
at the bottom of the Table\,\ref{tab:wd_ip}. Most likely, these objects have  low mass accretion rates.
The low mass accretion rate means that the local mass accretion rate is also low and, therefore, the shock and the PSR have a significant height, comparable with the WD radius. Thus, the models with a negligible shock height will underestimate the WD masses (see Eq.\,(\ref{eq:all})), and hence significant relative heights of the PSRs. This means that the WD masses we obtained in these systems are most likely also lower limits.
 
The closest low luminous eclipsing IP EX Hya requires separate consideration. 
The WD mass presented in Table~\ref{tab:wd_ip} was obtained using the common approach described above. However, the relative magnetospheric radius $r_{\rm m}$ in this system is significantly smaller than assumed here (0.75\,$r_{\rm C}$). Different authors give $r_{\rm m}$ values from 2.7 \citep{2014MNRAS.442.1123S} to 10 \citep{2024A&A...686A.304B}. The value $r_{\rm m}$=3.4 was obtained from simultaneously fitting the hard X-ray photon spectrum and the power density spectrum \citep{Suleimanov.etal:19}. 
However, the large PSR height can also be important, because the hard X-ray luminosity of the source is very low. 
We note that we assume that the WD surface temperatures are high (30\,kK). But the surface envelope temperatures in IPs with low mass accretion rates could be lower, down to 10\,kK  (see Eq.\,(\ref{eq:taccr})).
Therefore, it seems that we overestimated
the importance of the WD radii increase in low luminous IPs.

The average hard X-ray luminosity of GK Per, presented in the 105-month BAT catalog, is slightly higher than 10$^{33}$\,erg\,s$^{-1}$. However, we still put this source in the group of low luminous IP. This is because GK\,Per experiences dwarf-nova-like outbursts, during which the hard X-ray flux increases by an order of magnitude, approximately doubling the average flux. More accurate observations of GK\,Per, both during outbursts and in the quiescent state, were performed by NuSTAR. The case of this IP will be considered in a separate paper.

The accuracy and spectral resolution of BAT spectra are limited, leading to significant statistical uncertainties in the derived WD masses. Additionally, as a coded-mask instrument, BAT can be affected by source confusion, meaning the observed spectrum of an IP may be contaminated by hard X-ray emission from  nearby sources.
 An extreme example is IGR\,J08390$-$4833, situated nearby the Vela supernova remnant. As a result the WD mass determined using the BAT observations is much higher, 1.27$\pm$0.15\,$M_\odot$ \citep{Suleimanov.etal:19} than the WD mass found using the NuSTAR observation, 0.81$_{-0.11}^{+0.13}$\,$M_\odot$ \citep{Shaw.etal:20}. Comparison of the WD masses of other eleven IPs derived using BAT and NuSTAR observations demonstrated that the results may deviate with a spread of about 0.1\,$M_\odot$. This value is an additional systematic uncertainty which has to be added to the statistical errors presented in Table~\ref{tab:wd_ip}.   

\section{Discussion and conclusions}
\label{sect:discussion}

Hard X-ray spectroscopy is a powerful tool for determining the masses of WDs in IPs and polars. The hardness of the X-ray spectra of these sources is determined by the accretion-flow velocity above the shock and the structure of the post-shock region. The velocity above the shock is close to the free-fall velocity at the WD surface, and this fact is the basis of the hard X-ray spectroscopy method. However, the actual velocity could differ from the free-fall velocity for various reasons. We termed these reasons  the external systematic uncertainties and considered their influence on the WD mass determination. Uncertainties connected with the structure of PSRs were termed the inner systematic uncertainties, and these were not considered in the present paper.

Here we investigated the possible errors in determining the WD masses due to uncertainties in the plasma velocity above the shock. We considered uncertainties connected with the commonly used WD $M-R$ relation, the finite magnetospheric radii, the rotation of magnetospheres, the inclination of the impact accretion flow relative to the WD surface, and the finite shock height above the WD surface. Each of these factors leads to a decrease in the  accreting plasma velocity below the free-fall velocity. Therefore, using the free-fall velocity allows one to find only the lower limit of the WD mass. 

We found that the most important factor is the finite magnetospheric radius. The WD mass could be underestimated by up to 0.3\,$M_\odot$ if the relative magnetospheric radius $r_{\rm m}$ is about two. The importance of this factor was recognized quite a long time ago \citep[see, e.g.,][]{SRR:05, Suleimanov.etal:16}. However, the actual value of the magnetospheric radius for a given IP is unknown. The only thing known for certain is that the magnetospheric radius has to be smaller than the corotation radius $r_{\rm C}$. The equality of these radii ($r_{\rm m}=r_{\rm C}$) was used as a magnetospheric radius estimation in previous works \citep{Suleimanov.etal:19, Shaw.etal:20}. For some individual objects the magnetospheric radius can be estimated assuming that the frequency of the break in the IP power spectrum is equal to the Kepler frequency at the magnetospheric radius \citep{Revnivtsev.etal:09,Revnivtsev.etal:10, Suleimanov.etal:16}. Unfortunately, not all available IP light curves are of sufficient quality to obtain accurate power spectra. Moreover, this assumption can be not strictly correct \citep[see, e.g.,][]{2019MNRAS.486.4061M}. Therefore, the value of the magnetospheric radius is the main uncertainty for the WD mass determination of each individual IP.      

Two other factors that we considered for the first time also arise due to the finite magnetospheric radius. They are the centrifugal force originating from the rotation of the magnetosphere, and the inclination of the accretion flow relative to the WD surface. Both factors are less significant and their contribution is noticeable only for the case of a small magnetospheric radius, $r_{\rm m} < 10$ and $r_{\rm m} \approx r_{\rm C}$.

The uncertainty that does not depend on the magnetospheric radius is the used $M-R$ relation. In previous works, a relation for cold WDs was applied. 
Here we demonstrated that taking into account accretion-heated hydrogen-rich WD envelopes 
could be non-negligible to measure WD masses in IPs using the X-ray spectroscopy method.
The mass and chemical composition of the accreted envelopes in the CVs are poorly known. Therefore we used, as a first approximation, the models of thick ($\Delta M = 10^{-4}\,M_\odot$) hydrogen envelopes, presented by 
\citet{2001PASP..113..409F} and available in the database MWDD, to estimate the WD radii in IPs.
We demonstrated that taking into account WD radii with these envelopes leads to an increase in the measured WD masses by approximately 0.04\,$M_\odot$, assuming the envelope surface temperature is 30\,kK, which is typical for luminous IPs with a high mass accretion rate $\sim 10^{-9} M_\odot$\,yr$^{-1}$. A small difference between the average WD masses in IPs \citep[0.77--0.79\,$M_\odot$,][]{Shaw.etal:20, Suleimanov.etal:19} and CVs \citep[0.81--0.82\,$M_\odot$,][]{2022MNRAS.510.6110P,Zorotovic.etal:11} is probably explained by the use of the $M-R$ relation for cold WDs in IPs. 

To take into account the considered uncertainties we computed a new grid of hard X-ray spectra of PSRs on the surface of WDs with the envelope effective temperatures of 30\,kK, which is introduced into XSPEC. The fitting parameters are the WD mass (from 0.3 to 1.4\,$M_\odot$ with the step 0.02\,$M_\odot$) and the relative magnetospheric radius $r_{\rm m}$ from 1.5 to 60. It was assumed that the relative corotation radius $r_{\rm C}$ equals 10 if $r_{\rm m} <10$, and $r_{\rm C} = 1.25 r_{\rm m}$ otherwise. Corrections for the magnetospheric rotation and the accretion flow inclination relative to the WD surface were made based on these assumptions. A sufficiently high local mass accretion rate was assumed, and therefore the relative PSR heights were small and corresponding corrections were not taken into account.

We took a sample of 47 IPs observed by \emph{Swift}/BAT telescope and presented in the 105-month BAT catalog \citep{oh18}. Their hard X-ray spectra were fit by two spectral grids, the old one {\sc ipolar}, presented in XSPEC, and the new grid corresponding to the WD temperature of 30\,kK. For the IPs with known rotation periods we assumed that $r_{\rm m} = 0.75 r_{\rm C}$, and $r_{\rm m} = 60$ otherwise. The average WD mass obtained using the new models is 0.04\,$M_\odot$ higher than the corresponding value obtained with using the old grid, namely, 0.82$\pm$0.18 and 0.78$\pm$0.19\,$M_\odot$, respectively. This difference corresponds to the expections due to the introduction of a new $M-R$ relation for the WD with surface temperatures 30\,kK. Indeed, the obtained $r_{\rm m}$ for the investigated IPs are, as a rule, larger than 9--10, therefore, the other corrections not previously taken into account are insignificant.

There is no systematic shift between the average WD mass in IPs obtained here and the average WD mass obtained by optical methods \citep{Zorotovic.etal:11, 2022MNRAS.510.6110P}. It means that most of the WD masses in these IPs are accurate. Nevertheless, each individual mass should be considered a lower limit, because for some IPs the actual magnetospheric radius may turn out to be small.

We have additionally identified a separate subgroup of low luminous IPs. With a high degree of probability the local mass accretion rates on the WD surfaces in these systems are also low. Therefore, the relative PSR heights $h_{\rm sh}$ in these objects could be large. 
We demonstrate that this may lead to significant WD mass underestimations,  up to 0.15\,$M_\odot$ at $H_{\rm sh}/R \sim 0.4$. We note that three IPs with reliably determined WD masses (EX Hya, XY Ari, and DO Dra) are low luminous objects in hard X-rays. We are making the assumption that the found differences between WD masses obtained by using optical methods and the WD masses obtained by using hard X-ray spectroscopy arise due to the relatively high PSRs in the mentioned low luminous IPs. 
This fact is not yet taken into account with the necessary accuracy in the hard X-ray spectroscopy method applied to IPs
\footnote{Note, however, that in modern works on determining the WD masses in polars, tall accretion columns are considered with the necessary accuracy, see e.g., \citet{2025ApJ...987...53F}.}. But it does not mean that the masses of the high luminous IPs with the PSRs compressed against the WD surface are also strongly underestimated. 

It appears that the optical methods used by \citet{2021MNRAS.507.5805A,2023arXiv230408524A,2024arXiv240717562A} can  hardly be applied to luminous IPs.
Therefore, the hard X-ray spectroscopy is the only method for finding WD masses in these objects.

\begin{acknowledgements}
The work was supported by the German Research Foundation (DFG) grant WE\,1312/59-1 (VFS). VD thanks the Deutsches Zentrum for Luft- und Raumfahrt (DLR) and DFG for financial support. LD acknowledges funding from DFG - Projektnummer 549824807.
\end{acknowledgements}



\bibliographystyle{aa}
\bibliography{ip_lf} 

\begin{appendix} 
\section{Accretion on the magnetic dipole} \label{app:dip}

Dipole magnetic field lines are curved. Hence, the dynamics of the accretion flow is complicated and strongly depends on
the angle between the rotation axis and the magnetic dipole axis \citep[see, e.g.,][]{2023MNRAS.520.4315L}. Here we consider the accretion-flow dynamics according to the method described by \citet{2017MNRAS.467.1202M}.

Let us consider two limit cases, the magnetic dipole axis is parallel to the rotation axis (case 1), and the magnetic dipole axis is orthogonal to the rotation axis (case 2; see Figs.\,\ref{fig:dip1} and \ref{fig:dip2}). In both cases the radial coordinate $r$ is determined as
\be
        r = R_{\rm max}\,\cos^2 \lambda,
\ee
and the gravitational acceleration along the magnetic field line as
\be
       a_{\rm g} = \frac{GM}{r^2}\,\cos \chi
\ee
(see Fig.\,\ref{fig:dip1}). Here $R_{\rm max}$ is the maximum possible $r$ for a given magnetic field line at $\lambda=0$, and $\chi$ is the angle between the direction to the WD center and the tangent line to the magnetic field line in a given point. This angle is directly determined by $\lambda$
\be
      \tan \chi = \frac{1}{2} \tan^{-1}\lambda.
\ee
We note that in case 1, $R_{\rm max} = R_{\rm m}$ for all magnetic field lines along which the accreted matter flows. For case 2 this is true only for the magnetic field line which is tangent to the magnetospheric radius $R_{\rm m}$ (see Fig.\,\ref{fig:dip2}).

However, the projection of the centrifugal acceleration on the magnetic field line is different for both cases. The expression for case 1 is more complicated
\be
       a_{\rm c,1} = \omega_{\rm s}^2 r\, \cos \lambda \cos (\chi - \lambda)
\ee
than the expression for case 2
\be
       a_{\rm c,2} = \omega_{\rm s}^2 r\, \cos \chi.
\ee
As a result, the momentum equation can be written as
\be \label{eq:dip}
 {\rm v} \frac{d\vv}{d\lambda} = R_{\rm max}\,\Delta \lambda\, \left(a_{\rm g} - a_{\rm c,1,2}\right)
\ee
where $R_{\rm max}\,\Delta \lambda$ connects the angular variable $\lambda$ with the geometrical coordinate along the field line $x$
\be
        dx = R_{\rm max}\,\Delta \lambda\, d\lambda = R_{\rm max}\,\cos \lambda \left(4-3\cos^2 \lambda\right)^{1/2}\, d\lambda.
\ee
Equation (\ref{eq:dip}) can be reduced to the simplest form for case~2
\be \label{eq:dip2}
 \vv \frac{d\vv}{d\lambda} = R_{\rm max}\,\Delta \lambda\, \frac{GM}{r^2}\,\cos \chi \left(1 -\left(\frac{r}{R_{\rm C}}\right)^3 \right).
\ee
It is clear from this equation that the accretion is possible at any $r \le R_{\rm C}$. By integrating Eq.\,(\ref{eq:dip2})
we can show that Eq.\,(\ref{eq:rot}) is correct for case 2 for any values $r_{\rm m}$ and $r_{\rm C}$, if $r_{\rm m} \le r_{\rm C}$.

\begin{figure} 
\begin{center}  
\includegraphics[width=  0.9\columnwidth]{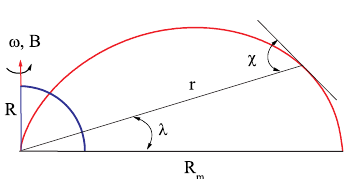}
\caption{\label{fig:dip1} 
Geometry of the accretion flow along the magnetic field line for case 1 (magnetic axis is parallel to rotation axis).
}
\end{center} 
\end{figure}

\begin{figure} 
\begin{center}  
\includegraphics[width=  0.9\columnwidth]{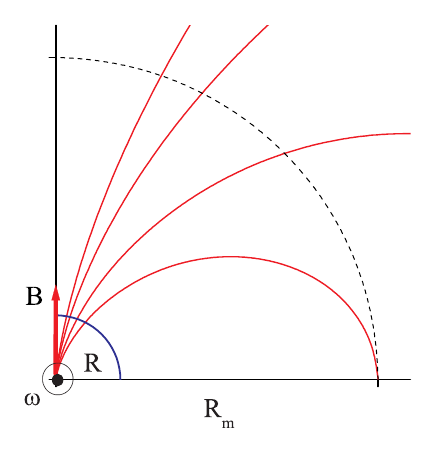}
\caption{\label{fig:dip2} 
Geometry of the accretion flow along the magnetic field line for case 2 (magnetic axis is orthogonal to rotation axis).
}
\end{center} 
\end{figure}

Case 1 is more complicated. Eq.\,(\ref{eq:dip}) becomes
\be \label{eq:dip1}
\vv \frac{d\vv}{d\lambda} = R_{\rm m}\,\Delta \lambda\, \frac{GM}{r^2}\,\left(\cos \chi \ - \omega_{\rm s}^2\,r\, \cos \lambda \cos (\chi - \lambda)\right).
\ee
and can be reduced at small $\lambda$ to
\be
       {\rm v} \frac{d{\rm v}}{d\lambda} = R_{\rm m}\,\Delta \lambda\, \frac{GM}{r^2}\,\left(1-\frac{3}{2}\left(\frac{R_{\rm m}}{R_{\rm C}}\right)^{3}\right).
\ee
It is clear that accretion is possible only at $R_{\rm m} \le (2/3)^{1/3}\,R_{\rm C} \approx 0.87\,R_{\rm C}$ in accordance with the result obtained by \citet{2023MNRAS.520.4315L}.

\section{Results of the fitting}
\label{app:tab3}

\begin{table*}
\caption{Observed data and derived WD masses for intermediate polars. 
 \label{tab:wd_ip} 
 }
 \begin{center}
\begin{tabular}{l|cccrrc}
 \hline \hline
Name &  $f_{14-195}^a$ & $d^b$ & $L_{14-195}$, & $P_{\rm s}$ & $r_{\rm m}$ & $M/M_\odot^c$ \\
  & $10^{-12}$\,erg\,cm$^{-2}$\,s$^{-1}$ & pc & $10^{32}$\,erg\,s$^{-1}$  &  sec &  &  \\ 
  \hline  \\
IGR J00234+6141  &	 $14.1(2.6)$	& $1557_{-72}^{+104}$	&	$41_{-9}^{+10}$   & 563.5 & 13.16&  0.99 $\pm$ 0.15  \\
 V709 Cas      &            $76.6(2.7)$	&	$725_{-9}^{+9}$	&	$48.2(2.1)$           & 312.8 & 8.08  & 0.94 $\pm$ 0.03  \\ 
 515 And  & $18.5(2.2)$	& $930_{-20}^{+17}$ &	$19.2(2.4)$     &   465.5 &   7.12 & 0.72  $\pm$ 0.08    \\ 
 IGR J04571+4527   &	$18.5(3.0)$	&	$1498_{-172}^{+185}$	&	$49_{-12}^{+16}$  &   1218.7  & 20.06  &0.94  $\pm$ 0.16  \\
 V1062 Tau            & $25(4)$	&	$1222_{-72}^{+88}$	&	$45_{-8}^{+9}$         & 3780.0  &31.93 & 0.77 $\pm$ 0.10  \\ 
 TV Col                 & $59.3(2.4)$	&	$502.8_{-3.5}^{+4}$	&	$18.0(8)$	    & 1910.0 & 21.30 & 0.80  $\pm$ 0.03   \\ 
 TX Col                    &       $12.3(1.6)$	&	$909_{-21}^{+18}$	&	$12.1_{-1.6}^{+1.7}$                    & 1911.0  &13.41  &0.56  $\pm$ 0.08  \\
 V405 Aur                &     $33.9(2.6)$	&	$658_{-8}^{+8}$	&	$17.6(1.4)$	                   & 545.5 & 8.95 & 0.80 $\pm$ 0.06  \\ 
 MU Cam          &      $15.6(1.8)$	&	$931_{-19}^{+23}$	&	$16.1_{-1.9}^{+2.0}$	  & 1187.3 &11.83& 0.65 $\pm$ 0.08 \\
 V647 Aur &  	$13.4(2.7)$	&	$1961_{-170}^{+192}$	&	$61_{-15}^{+18}$  & 932.9& 14.43 &0.85 $\pm$ 0.17 \\
 V418 Gem  &  $7.2(3.1)$	&	$3053_{-577}^{+866}$	&	$78_{-39}^{+63}$	& 480.7& 9.02& 0.84 $\pm$ 0.27 \\
 BG CMi            &   $23.6(2.5)$	&	$867_{-24}^{+26}$	&	$21.2_{-2.5}^{+2.7}$   &  913.5& 10.66& 0.69 $\pm$ 0.07  \\ 
 V667 Pup & $28.0(2.9)$	&	$1748_{-104}^{+131}$	&	$102_{-15}^{+18}$               & 512.4 &9.41 &0.84 $\pm$ 0.07 \\ 
 PQ Gem         &  $32.5(2.9)$	&	$734_{-14}^{+16}$	&	$20.9(2.0)$   & 833.4 &13.33 &0.85 $\pm$ 0.07  \\ 
  IGR J15094-6649     &    $24.4(3.0)$	&	$1091_{-23}^{+24}$	&	$35(4)$   & 809.4 & 12.40 &0.82 $\pm$ 0.08  \\ 
 NY Lup           &  $95.2(3.5)$	&	$1266_{-30}^{+35}$	&	$183_{-12}^{+13}$   & 693.0 &17.68 &1.07 $\pm$ 0.03  \\ 
  IGR J16500-3307   &  $22.5(2.4)$	&	$1091_{-52}^{+55}$	&	$32_{-4}^{+5}$  &  571.9 &  9.64& 0.81$\pm$ 0.08 \\
 IGR J16547-1916 &  $24.8(2.7)$	&	$990_{-43}^{+33}$	&	$29(4)$  &  546.7 & 9.76 &0.83 $\pm$ 0.09  \\ 
 V2400 Oph          &     $47.5(2.5)$	&	$700_{-11}^{+10}$	&	$27.8(1.7)$  &  927.7 &12.03 &0.75 $\pm$ 0.04  \\ 
 V2731 Oph & $68.7(3.2)$	&	$1938_{-138}^{+180}$   & 	$311_{-47}^{+61}$ & 128.0 &  6.47 & 1.13 $\pm$ 0.04  \\ 
 IGR J18173-2509    &    $14.6(2.3)$	&	$4690_{-1071}^{+1452}$	&	$373_{-182}^{+368}$   &  1663.4 & 13.86 & 0.62 $\pm$ 0.10 \\
 V1223 Sgr       &    $128.5(2.6)$	&	$561_{-8}^{+8}$	&	$48.4(1.6)$   &   745.6 &10.83 &0.77 $\pm$ 0.02  \\ 
 V2306 Cyg        &   $13.6(2.7)$	&	$1253_{-42}^{+51}$	&	$26_{-5}^{+6}$  & 1466.7 &14.79 &0.70 $\pm$ 0.10 \\
 V2069 Cyg        &    $17.8(2.6)$	&	$1155_{-39}^{+41}$	&	$28(5)$   &  743.2& 13.79& 0.91 $\pm$ 0.12 \\ 
 RX J2133.7+5107    & $55.0(2.9)$	&	$1372_{-35}^{+39}$	&	$124_{-9}^{+10}$    &  570.8 &12.19 &0.94 $\pm$ 0.04  \\ 
 FO Aqr      &   $52.9(3.2)$	&	$532_{-9}^{+7}$	&	$17.8(1.2)$   & 1254.0 & 12.78 &0.68 $\pm$ 0.03  \\
 SWIFT J0525.6+2416  &  $21(4)$	&	$1817_{-166}^{+183}$	&	$84_{-20}^{+27}$    &  226.3 & 6.23 & 0.91 $\pm$ 0.12 \\ 
 SWIFT J0614.0+1709 & $7.8(2.8)$	&	$1631_{-192}^{+270}$	&	$25_{-9}^{+12}$   & 1412.3 &  6.70 & 0.37 $\pm$ 0.25  \\
 SWIFT J0927.7-6945   &   $8.5(1.9)$	&	$1206_{-36}^{+34}$	&	$14.7_{-3.4}^{+3.5}$    & 1033.0 & 11.72 &0.70 $\pm$ 0.13  \\
 IGR J14257-6117     &    $12.5(3.2)$	&	$2082_{-336}^{+451}$	&	$64_{-26}^{+52}$   & 509.5 &  10.36 & 0.89 $\pm$ 0.22  \\
 SWIFT J1701.3-4304  & $10.6(3.0)$	&	$995_{-38}^{+41}$	&	$13(4)$  &  1859.0  & 25.35 & 0.91 $\pm$ 0.21   \\
IGR J18308-1232     &   $17.1(3.5)$	&	$2218_{-432}^{+512}$	&	$99_{-41}^{+87}$  & 1820.0 & 29.25 &1.00 $\pm$ 0.17  \\
 SWIFT J1832.5-0863   &  $19(4)$	&	$1481_{-419}^{+670}$	&	$48_{-30}^{+113}$  & 1549.0 &  39.82 &1.19 $\pm$ 0.20  \\
  IGR J18151-1052       &   $11.6(2.9)$	&	$4099_{-1683}^{+2655}$	&	$228_{-175}^{+526}$  & 390.5   &10.21 & 0.98 $\pm$ 0.25  \\
  SWIFT J0958.0-4208  &  $7.0(2.1)$	&	$1425_{-79}^{+89}$	&	$17_{-5}^{+6}$  &  7388.0  &37.13 &0.62 $\pm$  0.23  \\
 SWIFT J2006.4+3645   &   $14.4(3.2)$	&	$4459_{-1030}^{+1426}$	&	$337_{-193}^{+538}$  &  1466.7 & 27.90  &1.05 $\pm$  0.18  \\
 \hline  \\
 IGR J12123-5802      &  $13.3(2.7)$	&	$2736_{-338}^{+359}$	&	$118_{-34}^{+50}$	 &  $-$ &   60.0 &  $^{*}$0.94 $\pm$ 0.19  \\
 IGR J16167-4957  &  $22.0(2.7)$	&	$1543_{-127}^{+141}$	&	$63_{-12}^{+14}$  &  $-$ &  60.0 &  $^{*}$0.70 $\pm$ 0.08   \\
 V2487 Oph         &  $17.8(3.3)$	&	$6407_{-1634}^{+1646}$	&	$832_{-367}^{+644}$  &  $-$   &60.0 & $^{*}$1.23 $\pm$  0.15 \\ 
\hline  \\
EX Hya	 &	$26.3(2.5)$	&	$56.77_{-0.05}^{+0.05}$	&	$0.10(1)$	 & 4021.62 &    24.74 &     $^{**}$0.62 $\pm$  0.07 \\ 
  XY Ari                          & 30.5(2.2) & 308$\pm 40$ & 0.28(9)       & 206.3  &8.47  &$^{**}$1.11  $\pm$ 0.09 \\ 
  DO Dra              &  $17.6(2.1)$	&	$194.9_{-1.0}^{+1.1}$	&	$0.8(1)$  &  530.0 &8.03 &$^{**}$0.73 $\pm$ 0.09 \\
 V1025 Cen          &  $9.1(2.4)$	&	$196.0_{-3.0}^{+3.2}$	&	$0.4(1)$   &  2147.0 &14.68& $^{**}$0.57$\pm$ 0.16 \\
 YY Sex              &    	$9.8(3.2)$	&	$382_{-22}^{+31}$	&	$1.7(6)$	   & 1444.0 &  21.8 &  $^{**}$0.92 $\pm$ 0.32   \\
SWIFT J2113.5+5422 & $7.5(2.4)$	&	$612_{-44}^{+41}$	&	$3.3_{-1.1}^{+1.3}$ & 1265.6 & 10.19 & $^{**}$0.56 $\pm$ 0.13  \\
 AO Psc      &   $31.0(2.5)$	&	$462_{-5}^{+4}$	&	$7.9(7)$   &  805.2 &  8.12 &$^{**}$0.60 $\pm$ 0.04  \\ 
GK Per	  &	$61.0(2.7)$	&	$430_{-6}^{+7}$	&	$13.5_{-0.7}^{+0.8}$ & 351.0  &  6.89  &  $^{**}$0.80 $\pm$ 0.03 \\
 \hline 
\end{tabular}
\end{center}
Notes: a) Observed fluxes in 14--195\,keV band;  b) Distances from \citet{BJ21}, except for XY Ari, taken from \citet{2023arXiv230408524A}; c) an additional systematic uncertainty 0.1$M_\odot$ has to be added; 
$^*$ - lower limit due to possible smaller $r_{\rm m}$ ; $^{**}$ - lower limit due to possible large $h_{\rm sh}$.
\end{table*}

\end{appendix}

\end{document}